\documentclass[referee]{raa}
\usepackage{graphicx,times}             
\input{epsf.sty}                        
\input{psfig.sty} 

\begin{document}
 
\title{Photometric Study of W UMa Type Binaries in the Old Open Cluster Berkeley 39}
\subtitle{I. W UMa binaries in Be 39}

\author{K. Sriram
          , Y. Ravi Kiron 
           \and 
           P. Vivekananda Rao
          }
\institute{ Department of Astronomy, Osmania University,
              Hyderabad 500 07, India\\
              {\it astrosriram@yahoo.co.in}\\
       }

\abstract{The study of W UMa binary systems give a wealth of information about its nature as well as about its parent body (if any), like clusters. In this paper, we present the I passband photometric solutions of four W UMa  binaries in the open cluster Berkeley 39 using the latest version of W--D program. The result shows that, two binary systems are W-subtype W UMa binary systems and another two systems are H-subtype W UMa binary systems. No third body  is found in any of the four systems. We found a correlation between the period and mass-ratio as well as temperature and mass-ratio for the respective variables which is similar to the relationship between mass ratio and total mass of the contact binaries as shown by van't Veer (1996) and Li et al. (2008).     
}

\authorrunning{Sriram, Kiron \& Rao }            
\titlerunning{W UMa binaries in Be 39 }

\maketitle

\section{Introduction}
The old open cluster
Berkeley 39 is located in the direction of the galactic
anti-centre ($\alpha_{2000}$=7$^{h}$46$^{m}$48$^{s}$,
$\delta_{2000}$=-4$^{o}$40$^{'}$, l=223.5$^{o}$, b=10$^{o}$.1). 
A systematic survey of Berkeley 39 revealed 12 short period eclipsing binaries (Kaluzny et al. 1993) 
and later found 5 more variable stars in which one is W UMa type binary
system (Mazur et al. 1999).
The study of Color Magnitude Diagram (CMD) of Be 39 along with 
Padova stellar evolutionary models (Girardi et al. 1998) yields an age of 
around $6\pm{1}$ Gyr (Carraro 1998). Similar age constraint was proposed by 
Kassis et al. (1997) using the cluster CMD along with theoretical models by 
Bertelli et al. (1994). Applying theoretical and recent observational constrains, Percival \& Salaris (2003) have found an age of around 7.5$\pm$1.0 Gyr.
The de-reddened distance moduli for Be 39 is found to be 12.97$\pm$0.09 along 
with metallicity, [Fe/H]= - 0.15$\pm$0.09 (Percival \& Salaris 2003).

Since no photometric analysis is carried out for the eclipsing
binaries present in Be 39 cluster, we have analyzed the light curves of four eclipsing binary
systems V1,V4,V7 and V8 which have relatively high amplitudes. Similar
kind of study was carried out on the W UMa type binaries in open cluster NGC 6791 and Be 33 (Rukmini
\& Vivekananda Rao 2002, Rukmini et al. 2005). From the analysis performed on these systems, 
it was found that the mass-ratios 
were close to unity and in one case greater than unity (Rukmini
\& Vivekananda Rao 2002, Rukmini et al. 2005). A detail
spectroscopic observation was suggested to know the true nature of these objects. 
The aim of the present paper is to derive the photometric elements of
the selected four W UMa systems discovered in Be 39 using the latest W-D program and compare with the binaries studied in other clusters. Further it is also aimed to explored the possible correlation from the derived parameters. 

\section{Data and Analysis}
The observations were carried out by Kaluzny et al. 1993 at the Las Campanas
Observatory using the 1.0 m Swope telescope with 1k X 1k CCD in I passband
along with B and V passbands on the last day of the observation cycle. The
data was collected on the following 
six days in November (1990 Nov 29), December (1990 Dec 1, 2), January (1991 Jan 1, 29) and February (1991 Feb 26). We have used the I passband data published 
in that paper to obtain the photometric parameters of the respective systems (V1,V4,V7 and V8). Further observations of the cluster were carried by the same group during 1992/1993 and they discovered new variable stars along with previous ones. Out of the total 17 variable stars discovered in Be 39 12 variables belongs to EW type, 3 of EA type, 1 is $\delta$ Scuti and 1 is EB type. Kaluzny et al. (1993) have found that, four variables are blue stragglers which mimics the light curves of W UMa type binaries (V1, V2, V10, V12).

The observed data of selected variables in cluster Be 39 are plotted  
to obtain the respective light curves (see Fig 1). All of them have amplitude greater than 0.4.
The phase of the variables were calculated using the accurate periods published in Mazur et al. (1999).  

\section{Photometric Solutions} 

The photometric solutions were obtained by using latest Wilson-Devinney
program with an option of non-linear limb darkening via a square root
law along with many other features (\cite{wd71, van03}). The visual inspection 
of these variables resembles W UMa type nature variability 
and hence mode 3 was used in the program. The parameters adopted in the
solutions are as follows: the effective temperature for star 1 is taken on the basis of un-reddened B-V values
published by Mazur et al. (1999) and using the Allen's table (2000). 
The secondary component
effective temperature ($T_{e,c}$) was assumed to be equal to primary component's temperature. The
values of limb darkening coefficients, x$_{h}$ \& x$_{c}$ = 0.80 for
I band were taken from AlNaimy (1978). The gravity-darkening coefficients
G$_{h}$ \& G$_{c}$ for I passband was set to 0.32 and the values of albedo A$_{h}$ \&
A$_{c}$ was fixed at 0.5 which is appropriate for the convective nature
of the component stars.  

Since no spectroscopic observations are available for the variables, We had to constrain the most important parameter 
mass-ratio q=m2/m1. In order to find the best value of mass-ratio, we executed the code for various assumed values of mass-ratio keeping in the mind that the surface potential ($\Omega_{h}$=$\Omega_{c}$)
also changes with respect to mass-ratio. Initially all the parameters were fixed as 
mentioned above, except q and i which are adjustable parameters. Further both these parameters are not adjusted simultaneously. Several runs of the DC program was executed to obtain the
minimum $\Sigma$W(O-C)$^{2}$. After performing several computational runs, we found 
the best mass-ratio value for the respective light curves. The Table~\ref{tab 1} show the values of $\Sigma$ for different combinations of q and i.  

\begin{figure}
\includegraphics[height=15cm,width=5cm, angle=270]{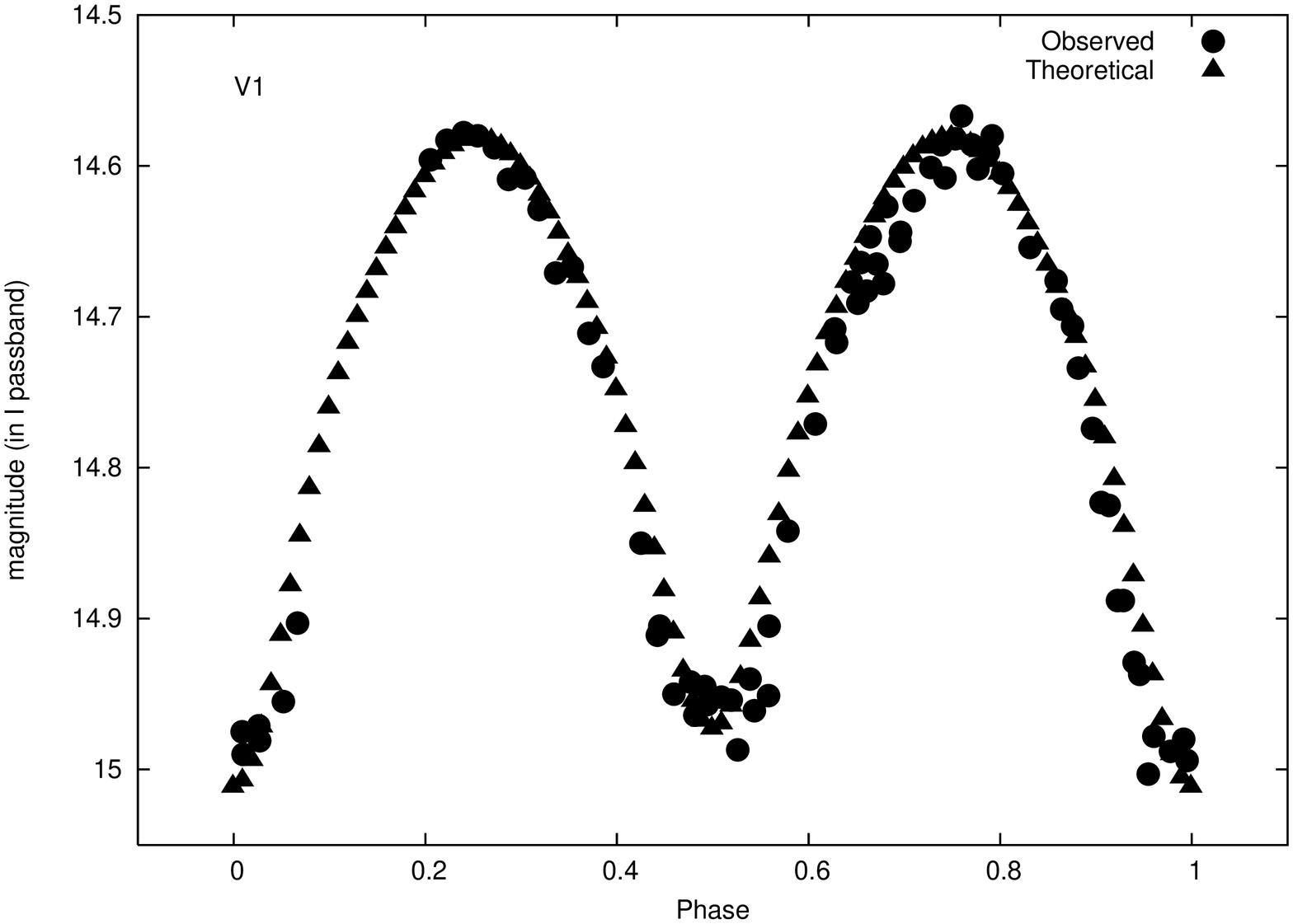}\\
\includegraphics[height=15cm,width=5cm, angle=270]{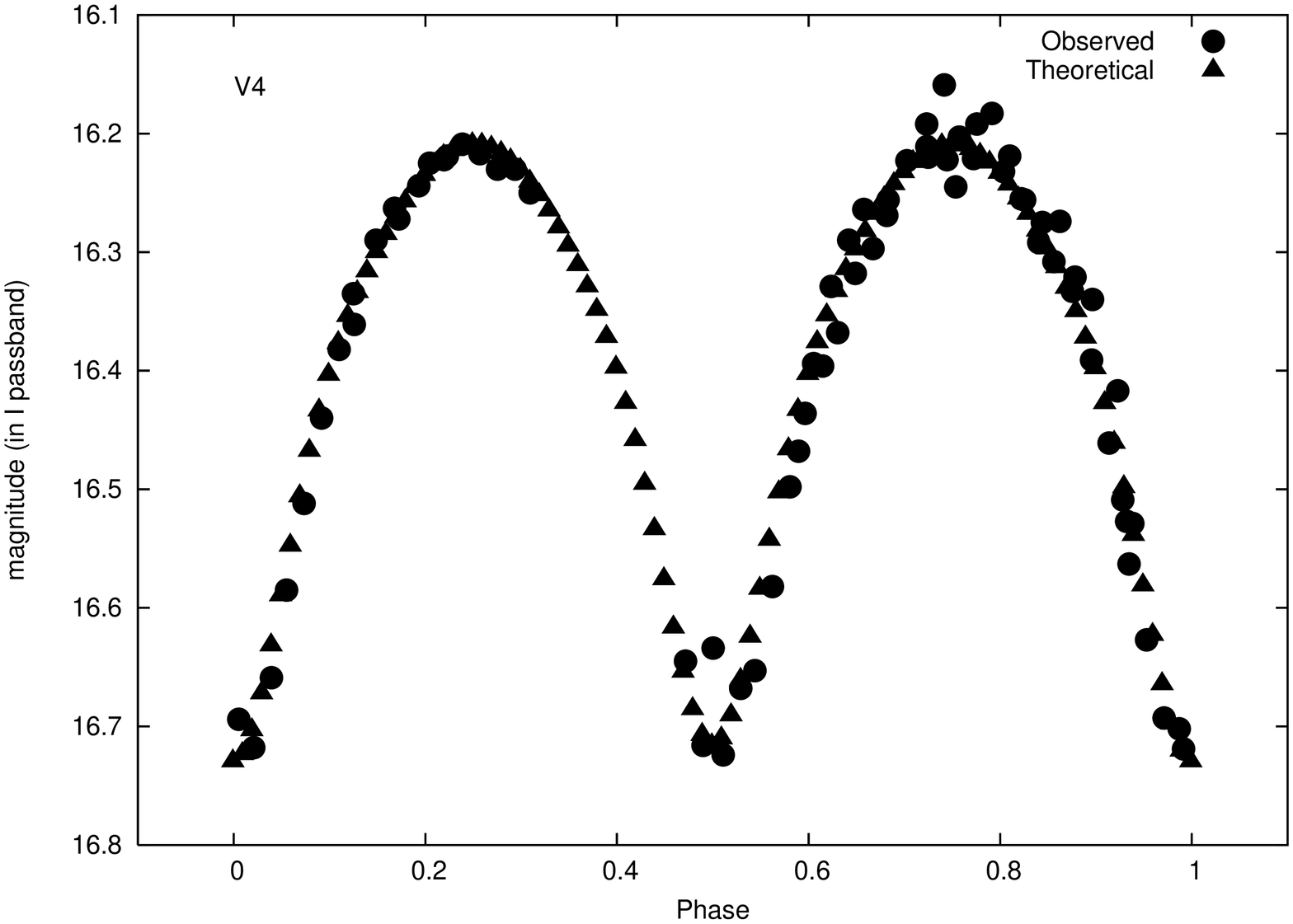}\\
\includegraphics[height=15cm,width=5cm, angle=270]{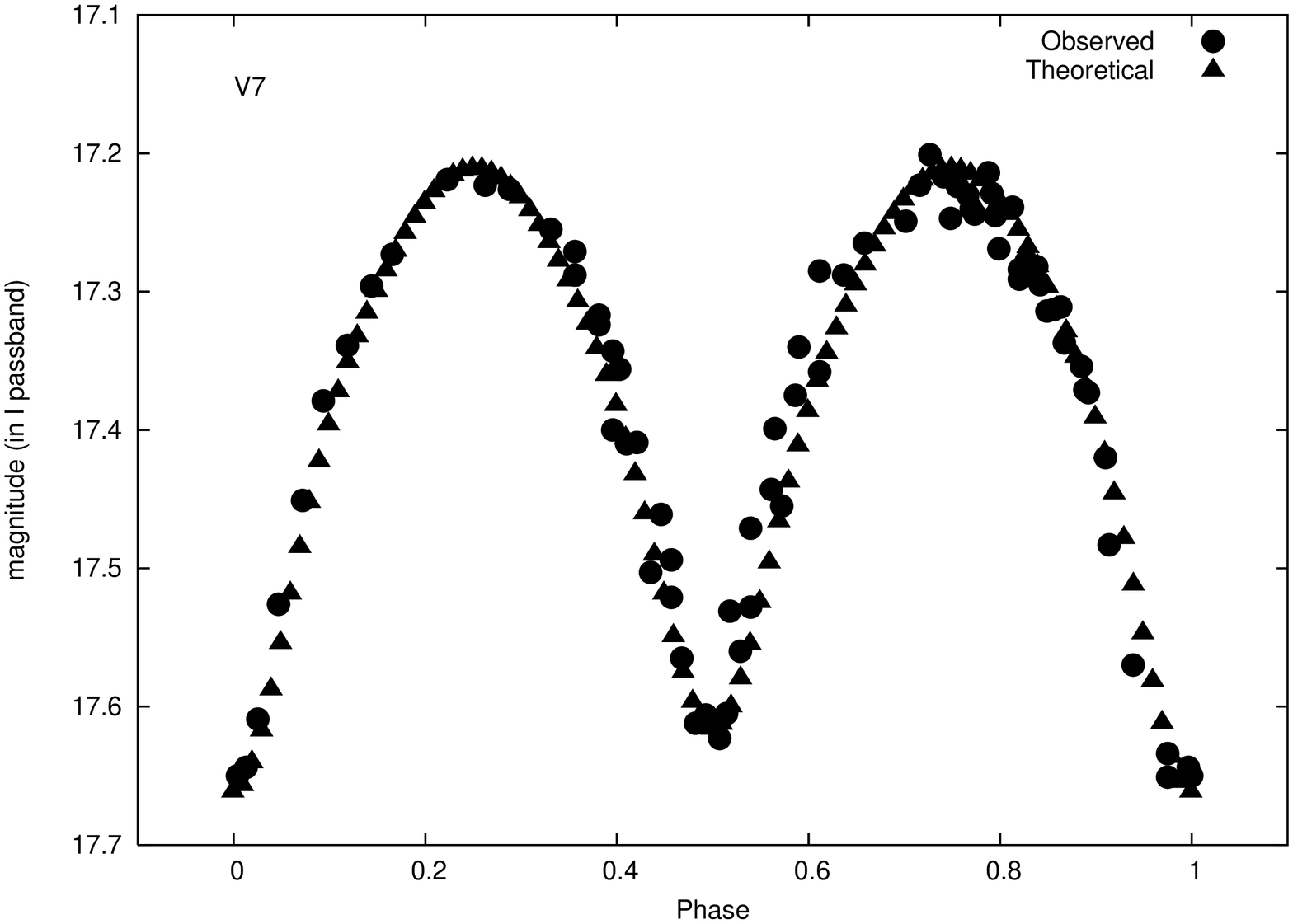}\\
\includegraphics[height=15cm,width=5cm, angle=270]{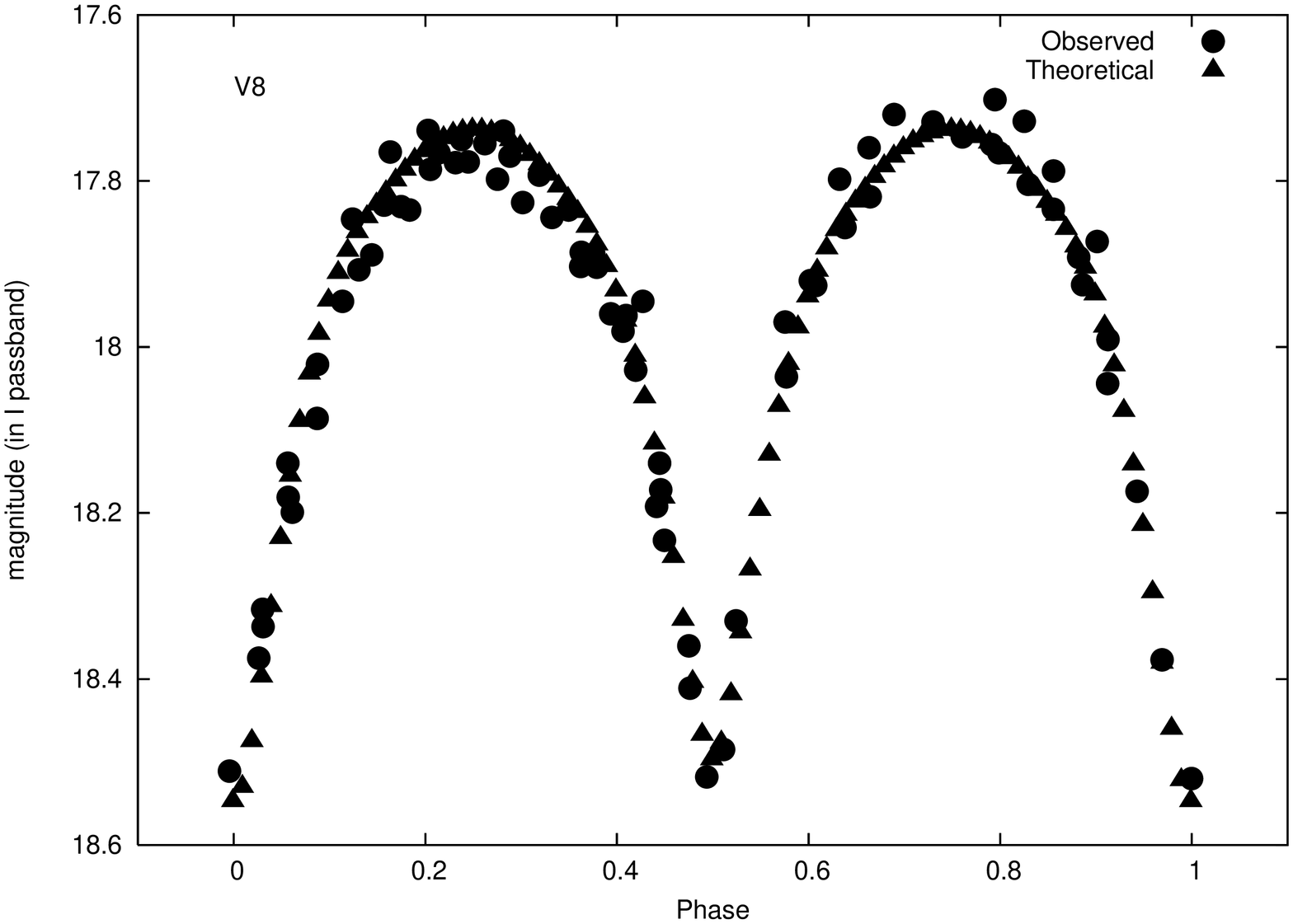}\\
\caption{The figure shows the best fit to the I passband light curve of the four variables present in Berkeley 39. The filled circles represent the observed data and filled triangles represent the theoretical points.\label{fig1}} 
\end{figure}

After obtaining the best set of q and i, the DC program was executed to obtain the values of the following parameters i.e. inclination (i),
secondary component temperature ($T_{e,c}$), surface potential ($\Omega$) and monochromatic 
luminosity of primary component (L$_{h}$), third light parameter (L$_{3}$) and limb darkening coefficients (x$_{h}$ \& x$_{c}$) . The program was executed till the 
sum of residual $\Sigma$W(O-C)$^{2}$ attained a minimum. The result of the final analysis 
are shown in Table~\ref{Table 2}. Using the final parameters given in Table 2, theoretical light curves were computed using the LC program of Wilson-Devinney and the fits are shown in Fig~\ref{fig1}. The quality of the fits were checked by performing chi square ($\chi^{2}$) test on the $\Sigma W (O-C)^{2}$ values obtained and the confidence level were found to be about 95\% for all the four variables.



\section{Discussion and Result}

The study of W UMa variable systems present in clusters are important in-order to
understand the underlying physics of there formation and also provides some information of the
parent cluster. In this paper, we have analyzed four eclipsing variables present in old open cluster Berkeley 39.
We used the latest W-D program to derive the photometric solutions. The result on each of the system is discussed below. 

\subsection{Variable V1}
The proper sinusoidal variation of the light curve indicate towards W UMa type binary nature. The period
of the system is around $\sim$0$^{d}$.4052 and B-V=0.66 which corresponds to G5 spectral 
type (Kaluzny et al. 1993, Mazur et al. 1999). The maxima and minima
 are not the same in the I passband and the difference is more clear in V passband. The best combination of q and i came out
to be 1.8 \& $\sim$ 68$^{o}$ respectively. The computed values of the effective temperature of 
the primary and secondary show a difference of 200 K.

\subsection{Variable V4}
The observed and theoretical light curve suggest that the variable is W type W UMa binary system. The period of the 
system is around  $\sim$0$^{d}$.3813 and the B-V=0.66 value correspond to G5 spectral type. The light curve 
maxima and minima are equal in I and V passbands and the effective temperatures of the primary and secondary component
are nearly same. The best combination of q and i is 1.31 \&  $\sim $73$^{o}$ respectively.  

\subsection{Variable V7}
The period of the W UMa binary system is around $\sim$0$^{d}$.2780 and the value of B-V=0.82 suggests 
K0 spectral type. The maxima and minima of the light curve are not same in I and V passbands and its difference is 
comparatively more in the V passband. The best combination of q and i came out be 1.38 and $\sim$ 67$^{o}$ respectively.
  
\subsection{Variable V8}
The period of the variable was found to be $\sim$0.$^{d}$2288 and value of B-V=1.02 indicates K2-K3 spectral type. The maxima and minima of the light curve are equal in both I and V passbands. No temperature difference is found 
between primary and secondary components. This variable has a high inclination angle $\sim$ 83$^{o}$ compared to other variables mentioned above. The mass -ratio is found to be $\sim$1.47.\\

Here we have taken the inverse mass-ratio values for further discussion. The inverse mass-ratio (q*) values for the studied variables are
q*=0.55 for V1, q*=0.76 for V4, q*=0.72 for V7, q*=0.68 for V8. The variables V4 \& V7 belong to H type W UMa binary systems, as there mass-ratio values are greater than 0.72 according to Csizmadia \& Klagyivik (2004). The H-subtype classification of contact binaries is the result of luminosity ratio vs energy transfer parameter study and it was found that the energy transfer rate is less efficient when compared to other contact binary stars (Csizmadia \& Klagyivik 2004). Qian (2001) has studied 30 W type W UMa binary systems and found that if mass-ratio q$>$0.4 then the period increases and if q$<$0.4 the period decreases with respect to mass-ratio. This result is explained by the contact configuration of the systems is due to variable angular momentum loss (AML) mechanism and thermal relaxation oscillation (TRO) cycle (Qian 2001). Since the result is based on a small sample, it cannot be extended to other W UMa contact binary systems and if it is a real effect then theoretical interpretation is needed.\\

In our study of variables in the open cluster Be 39, four W UMa systems (two W-subtype and H-subtype) have mass--ratio greater than 0.4 suggesting that they are in period increasing phase according to Qian (2001). It was found that for all the W-subtype W UMa systems the period increases along with temperature (Rucinski 1998). Our results indicate that mass-ratio is increasing along with the period of the respective binary systems. We have looked for the light curve solutions of W UMa type contact binaries (Csizmadia \& Klagyivik 2004) whose periods are small ($<$0.$^{d}$28) and posses relatively high mass-ratio value (q$>$0.55) and we found that there are total 7 W UMa type systems out of which 4 are H-subtype and remaining are W-subtype. The V1 variable in NGC 6791 (Rukmini et al. 2005) also falls in this category (period 0.$^{d}$2677 \&  mass-ratio, q*=0.83). For all these W UMa systems, the mass-ratio increases along with the respective period (Fig 2). The derived photometric mass-ratio of three variables i.e V4, V7 \& V8 (V1 is not the member of the cluster) and CMDs published by Mazur et al. (1999) show a correlation between temperature and mass-ratio viz. the temperature increases along with the mass ratio for the respective variables. The fractional change in the mass-ratio is about $\sim$ 11 percent for a fractional change of $\sim$ 16 percent in temperature of the primary component of the respective variable. The mass ratio q changes from 0.68 to 0.76 for a temperature change of 4652 K - 5507 K, which is consistent with the study carried out by Rucinski (1998) (see Fig. 4, Fig. 5 \& Fig. 6 in Rucinski (1998)). The relationship between mass ratio and orbital period and a relationship between the effective temperature and mass ratio for variables in cluster Be 39 are similar to the relationship between mass ratio and total mass of the contact binaries as shown by van't Veer (1996) and Li et al. (2008). To confirm the photometric mass ratio of the variables in Be 39, spectroscopic studies are needed.


\begin{acknowledgements}
     The authors acknowledge an anonymous referee for the useful comments. K.S is supported by UGC through {\it Research Fellowship in Science for Meritorious Student} scheme. 
\end{acknowledgements} 

\begin{figure}
\includegraphics[height=15cm,width=10cm, angle=270]{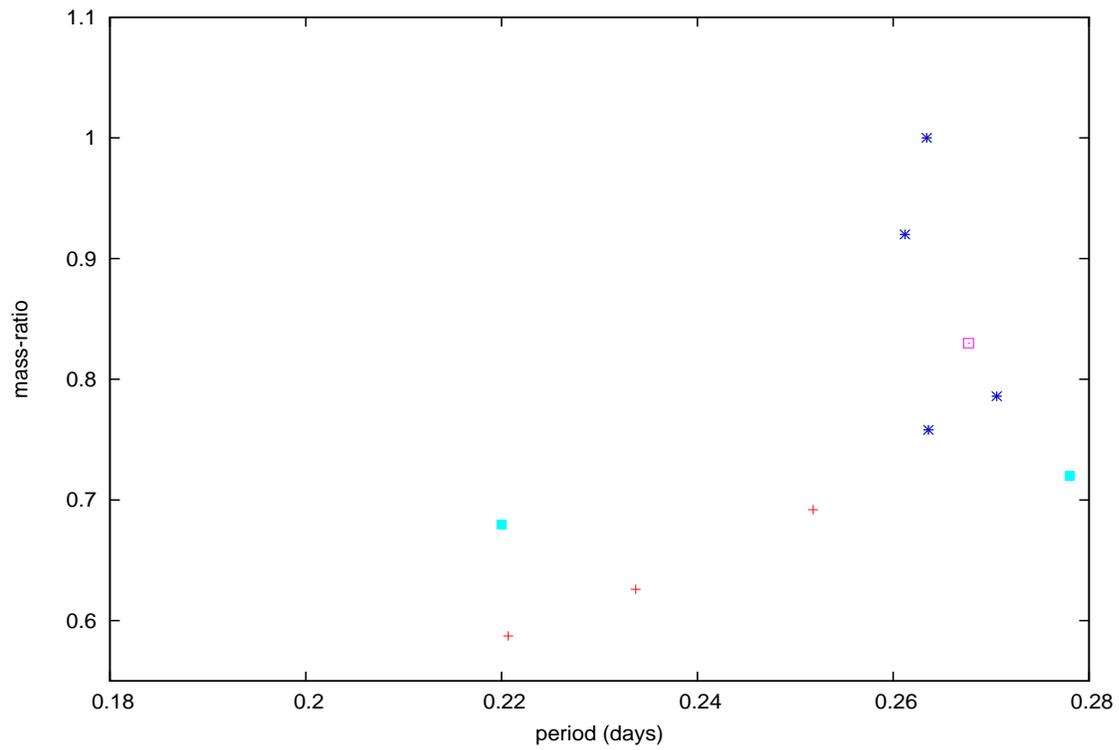}\\
\caption{The figure shows increasing trend of period vs mass-ratio. The plus symbol represents the W-subtype, star symbol represents the H-subtype, dot box represent the V1 variable in NGC 6791 and filled box symbol represent the variables V7 \& V8 in Be 39. \label{fig2}} 
\end{figure}


{\begin{table*}

\begin{minipage}[t]{\columnwidth}
\caption{The values of $\Sigma$ for various combinations of q and i, for V1, V4,V7 and V8  
variables (see the text). \label{tab 1}}
\renewcommand{\footnoterule}{}
\centering
\begin{tabular}{ccccccccc}
\hline
\multicolumn{3}{c}{ V1\footnote{Variable name}} & \multicolumn{2}{c}{ V4} &
\multicolumn{2}{c}{V7} & \multicolumn{2}{c}{ V8} \\ 
q&i  & $\Sigma$ & i & $\Sigma$ & i &$\Sigma$&  i & $\Sigma$  \\
\hline
0.5     &01.22 &0.084  		&05.35 &0.099            &10.76 &0.043    	   &10.25&0.107\\
0.6     &03.22&0.074   		&14.76 &0.063            &22.023 &0.035            &27.49 &0.094\\
0.7     &07.88&0.065            &30.71 &0.051            &33.54 &0.027             &37.87  &0.081\\
0.8     &11.32&0.055            &41.84 &0.039            &42.66 &0.021 		   &47.30  &0.069\\
0.9     &21.61&0.046            &51.98 &0.028            &50.64 &0.014  	   &56.62 &0.055\\
1.0     &31.38&0.038            &59.68 &0.018            &56.862 &0.008 	   &63.10 &0.041\\
1.1     &38.90 &0.031           &65.04 &0.009 		 &61.891 &0.004		   &68.31 &0.028\\
1.2     &45.83  & 0.024         &69.20 &0.003 		 & 65.77 &0.0017 	   &72.10 &0.018\\
1.3     &51.70  &  0.018        & 72.66 & 0.001   	 &69.08 &0.002  	   & 75.30 &0.010\\
1.4     &57.62  & 0.010         &75.67 &0.003   	 &71.98 & 0.005 	   &78.22 &0.0052\\
1.5    & 60.14& 0.007           &78.48 &0.009  		 &74.16 &0.011 		   & 80.98&0.0035\\
1.6     &63.25  &0.004          &81.19 &0.018     	 &  -  & -		   &84.125 &0.005\\
1.7     &65.83 &0.0018          &83.11 &0.028      	 & -   & -		   &87.90 &0.010\\
1.8     &68.194&0.0017    &84.26 &0.120            & -   & -		   &92.12 &0.017\\
1.9   &70.23   &0.0032          &  -    & -              & -   & -                 & -   &-\\
2.0   &72.42   &0.0065          &  -     & -             & -   & -                 & -   &-\\ 
\hline
\end{tabular}
\end{minipage}
\end{table*}}

\clearpage

{\begin{table*}[1]
\begin{minipage}[t]{\columnwidth}
\caption[solutions]{The Photometric elements obtained for four variables by using W-D method.}
\label{Table 2}
\renewcommand{\footnoterule}{}
\centering
\begin{tabular}{ccccccc}
\hline
\hline
Element & & V1 & V4 & V7 & V8\\ \\
\hline
\hline
Period (days)&&0.4052&0.3813&0.2780&0.2288\\
$T_{e,h}\footnote{Fixed parameters} K$ & &5560                  & 5560 			& 5150			&4700 \\ 
$T_{e,c} K$ & & 5291$\pm$84         & 5507$\pm$77 		& 4900$\pm$84 		&4652$\pm$50  \\
q$^{a}$ & & 1.80                       & 1.31 			& 1.38 		         &  1.47\\ 
q*\footnote{inverse mass-ratio}&&0.55&0.76&0.72&0.68\\
i$^{o}$ & & 68.24$\pm$0.50            & 72.97$\pm$0.46 		& 66.68$\pm$0.48	 &83.07$\pm$0.92 \\ 
$\Omega$ &  &4.7426$\pm$0.0038& 4.2055$\pm$0.0024        &4.2043$\pm$0.0037       &4.5868$\pm$0.0043\\
f\footnote{Fill-out factor} & &.1338&.0315&0.2997&0.1050\\

$r_{h}$ & pole & 0.3187$\pm$0.0032     & 0.3332$\pm$0.0031 	& 0.3387$\pm$0.0021 	&  0.3147$\pm$0.0020 \\
        & point & 0.4152$\pm$0.0041     & 0.4568$\pm$0.0043 	& 0.4528$\pm$0.0041	& 0.3930$\pm$0.0052  \\
        & side & 0.3328$\pm$0.0022 	& 0.3489$\pm$0.0022 	& 0.3547$\pm$0.0025 	&  0.3279$\pm$0.0032 \\
        & back & 0.3627$\pm$0.0023 	& 0.3803$\pm$0.0030	& 0.3854$\pm$0.0030 	& 0.3550$\pm$0.0035  \\

$r_{c}$ & pole & 0.3862$\pm$0.0022 	& 0.3782$\pm$0.0021 	& 0.3705$\pm$0.0021 	& 0.3789$\pm$0.0038 \\
 	& point & 0.4980$\pm$0.0041 	& 0.5123$\pm$0.0042 	& 0.4919$\pm$0.0033	&  0.4710$\pm$0.0062 \\
 	& side & 0.4073$\pm$0.0042  	& 0.3986$\pm$0.0030 	& 0.3898$\pm$0.0028 	&  0.3984$\pm$0.0042 \\
	& back & 0.4351$\pm$0.0050 	& 0.4284$\pm$0.0032 	& 0.4193$\pm$0.0031 	& 0.4241$\pm$0.0041  \\

$L_{h}$\footnote{In units of total system at phase 0.25}& & 0.4565$\pm$0.0123 & 0.4448$\pm$0.0098  & 0.4991$\pm$0.0153 &  0.4179$\pm$0.0098 \\

$L_{c}^{d}$& & 0.5435 	      & 0.5551 		   & 0.5008   	&  0.5821 \\ 

$L_{3}$ & & $0.0$ & $0.0$ & $0.0$  &$0.0$\\ 

$x_{h}$ & & 0.80$\pm$0.05 			& 0.80$\pm$0.05 		&  0.80$\pm$0.05 	& 0.80$\pm$0.05 \\ 

$x_{c}$ & & 0.80$\pm$0.05 			& 0.80$\pm$0.05			&  0.80$\pm$0.05 	        &0.80$\pm$0.05  \\
$\Sigma$ & &0.00132                     &0.00121                &0.00146                &0.00285\\
Spectral type & &G5                    &G5                     &K0                     &K2-K3\\
A$_{h}^{a}$ & & 0.5 & 0.5 & 0.5  & 0.5&\\ 
A$_{c}^{a}$ & & 0.5 & 0.5 & 0.5  & 0.5\\ 
\hline
\hline
\end{tabular} 
\end{minipage}
\end{table*}}

\end{document}